\documentclass[conference]{IEEEtran}
\IEEEoverridecommandlockouts
\usepackage{amsmath,amssymb,amsfonts}
\usepackage{algorithmic}
\usepackage{graphicx}
\usepackage{textcomp}
\usepackage[table]{xcolor}

\usepackage{verbatim}
\usepackage{amsmath}
\usepackage{array}
\usepackage{dcolumn}
\usepackage{bm}
\usepackage{setspace}
\usepackage{amssymb}
\usepackage{braket}
\usepackage{graphicx}
\usepackage[caption=false]{subfig}

\usepackage{ragged2e}
\usepackage{multirow}

\usepackage{verbatim} 
\usepackage{mathtools}
\usepackage{enumitem}
  {\list{}{\leftmargin=#1\rightmargin=#1}\item[]}%
  {\endlist}
  
\begin{document}
\bstctlcite{IEEEexample:BSTcontrol}
\author{\IEEEauthorblockN{
Charlene Yang 
}

\IEEEauthorblockA{
National Energy Research Scientific Computing Center (NERSC) \\
Lawrence Berkeley National Laboratory (LBNL) \\ Berkeley, California 94720, USA \\
}
\IEEEauthorblockN{
{
cjyang@lbl.gov 
}
}
}
\title{8 Steps to 3.7 TFLOP/s on NVIDIA V100 GPU: Roofline Analysis and Other Tricks}

\maketitle

\begin{abstract}
Performance optimization can be a daunting task especially as the hardware architecture becomes more and more complex. 
This paper takes a kernel from the Materials Science code BerkeleyGW, and demonstrates a few performance analysis and optimization techniques.
Despite challenges such as high register usage, low occupancy, complex data access patterns, and the existence of several long-latency instructions, we have achieved 3.7 TFLOP/s of double-precision performance on an NVIDIA V100 GPU, with 8 optimization steps. 
This is 55\% of the theoretical peak, 6.7 TFLOP/s, at nominal frequency 1312 MHz, and 70\% of the more customized peak based on our  58\% FMA ratio, 5.3 TFLOP/s.
An array of techniques used to analyze this OpenACC kernel and optimize its performance are shown, including the use of hierarchical Roofline performance model and the performance tool Nsight Compute. 
This kernel exhibits computational characteristics that are commonly seen in many high-performance computing (HPC) applications, and are expected to be very helpful to a general audience of HPC developers and computational scientists, as they pursue more performance on NVIDIA GPUs. 

\end{abstract}


\section{Introduction}

The Roofline performance model \cite{CACM09_Roofline1} has gained a lot of popularity in recent years for performance characterization, analysis and optimization in high-performance computing (HPC). 
It provides useful insights to machine characteristics, bottlenecks of application performance, and performance optimization strategies. 
Thanks to the community's research interest on Roofline, the model has been expanded to characterize the full memory hierarchy, instead of focusing on the highest level cache/memory only.
Methodologies for collecting Roofline data have been proposed on Intel CPUs \cite{koskela2018novel} and NVIDIA GPUs \cite{yang2019hierarchical,metho,yang2020howto}, and they have been integrated into production tools Intel Advisor \cite{advisor} and NVIDIA Nsight Compute \cite{ncu} respectively.
The hierarchical Roofline helps understand cache reuse and data locality, providing even more insights into how efficiently the code is using the memory subsystem.  

To facilitate the Roofline study, a range of tools have sprung to life, for more accurate machine characterization such as the Empirical Roofline Toolkit (ERT) \cite{ert,yang2018empirical}, and for more streamlined methods to collect Roofline performance data using open-source tools or workflows \cite{nerscroofline,yang2018toward,madsen2020timemory,yang2019hierarchical,yang2018toward}.
Other than tools development, there are also many studies on the application of the Roofline model in both traditional HPC \cite{Doerfler,yang2019hierarchical,del2020accelerating,gayatri2018case} and the new, emerging field of Machine Learning \cite{yang2019hierarchical,wang2020pmbs,wang2020dlonsc}, and the extension and refinement of the model, such as instruction Roofline \cite{ding2019instruction}, Roofline scaling trajectories \cite{ibrahim2019performance}, performance portability based on Roofline \cite{yang2018empirical}, and power and energy Roofline \cite{powerroofline,alexpowerroofline}.


In this paper, we focus on a General Plasmon Pole (GPP) kernel from the BerkeleyGW code and discuss what difficulties scientific HPC codes are usually faced with and what performance analysis and optimization techniques can be employed to achieve good performance on NVIDIA GPUs.
To that end, we will use the Roofline model for high-level performance analysis \cite{yang2019hierarchical}, and the performance tool Nsight Compute \cite{nsight_compute} for more detailed performance data collection.
The 8 optimization steps taken to speed up the GPP kernel by 3 times include, replacing long-latency instructions with shorter ones, rearranging loops to gain arithmetic intensity, reducing branching, and cache blocking. 
With these steps, we have achieved 3.7 TFLOP/s double precision performance on an NVIDIA V100 GPU, which is 55\% of the theoretical peak 6.7 TFLOP/s at nominal frequency 1312 MHz, and 70\% of the more customized peak based on our  58\% FMA ratio, 5.3 TFLOP/s.
This is despite the challenges such as high register usage, low occupancy, complex data access patterns, and the existence of several long-latency instructions. 
An array of techniques used to analyze this OpenACC kernel and optimize its performance are demonstrated, and due to the commonality of computational characteristics this kernel shares with other HPC application, these techniques are expected to be very helpful to a large segment of audience in HPC, as computational scientists and programmers embark on their quest of performance optimization on NVIDIA GPUs.



The rest of the paper is organized as follows. 
Section II will describe the GPP kernel and its implementation in detail, Roofline data collection methodology used in this paper, and the machine configuration.
Section III will then discuss the 8 optimization steps taken to improve the kernel's performance from 2.3 TFLOP/s to 3.7 TFLOP/s on an NVIDIA V100 GPU.
A combination of hierarchical Roofline analysis and the performance tool Nsight Compute is employed, and a more customized Roofline ceiling based on our FMA ratio is presented.
Finally, Section IV will draw conclusions on lessons learned through this study, which will be helpful to many GPU developers and computational scientists. 



\section{Application and Methodology}

\subsection{General Plasmon Pole (GPP) Kernel\label{sec:gpp}}

The GPP kernel~\cite{metho} is abstracted from the Sigma module of the Materials Science code BerkeleyGW~\cite{BGW2}, commonly used for  self-energy calculations in electronic structure studies.
It is written in Fortran and parallelized with OpenACC, and 
the computation involved in this kernel represents work that typically is performed on an individual MPI task in a much larger calculation.
The computation is tensor-contraction like, and several pre-calculated complex double precision arrays are multiplied and summed over certain dimensions and collapsed into a small vector. 

We use two benchmark systems for this study, Si-214 and Si-510, respectively, a silicon system with 214 atoms and 510 atoms.
The pseudo code of the kernel is shown below, and the magnitudes for different loops are listed for the Si-214 system.
The Si-510 system is 3 to 4 times larger on the \texttt{band}, \texttt{igp}, and \texttt{ig} loops than Si-214. 

{\footnotesize
\begin{verbatim}
  start timer
  do band = 1, nbands # O(1000)  
   do igp = 1, ngpown # O(1000)  
    do ig = 1, ncouls # O(10000)  
     do iw = 1, nw    # small, nw=2 
       load wtilde_array(ig,igp)  
       load aqsntemp(ig,band)  
       load eps(ig,igp)  
       compute wdiff, delw, sch_array, ssx_array
       reduce on achtemp(iw), asxtemp(iw)
  end timer
\end{verbatim}
}

Some computational characteristics of the GPP kernel are, 1) there is abundant parallelism, enough to saturate the GPU for efficient utilization, 
2) most arrays in this kernel are in double precision, either floats or complex numbers,
3) dimensions of the arrays encompass almost all combinations of indices \texttt{band}, \texttt{igp} and \texttt{ig}, possibly posing a challenge in ensuring memory coalescence between threads in a warp, and the effectiveness of cache blocking,
4) among many multiplications, additions, and fused-multiplication and additions (FMAs), there are a few divides and \texttt{abs()} instructions, which present a longer latency and could potentially hurt the performance,
5) the reduction runs across all three loops \texttt{band}, \texttt{igp}, and \texttt{ig}, and accumulates results to the \texttt{iw}-th element of arrays \texttt{achtemp} and \texttt{asxtemp}.
At the time of writing, OpenACC does not support array reductions, so some alternative needs to be sought. 
Luckily the loop count $nw$ is very small and fixed so scalar reductions can be used for each $iw$ explicitly.
In PGI's implementation of OpenACC, scalar reductions are usually done in a separate kernel after the main computational kernel, with much shorter computation and much fewer threads. There will be two kernels appearing in the Nsight Compute profile, however, we will only focus on the compute kernel for the bulk of this performance study. The runtime and throughput calculation, though, (Tab.~\ref{tab:perf}, and Fig.~\ref{fig:r0}, Fig.~\ref{fig:r1}, Fig.~\ref{fig:r3} and Fig.~\ref{fig:r5}), will still be for the entire calculation, as indicated by the timers in the pseudo code above.  


\subsection{Roofline Data Collection Methodology\label{sec:metho}}

The Roofline model \cite{CACM09_Roofline1} characterizes application performance based on two quantities, arithmetic intensity (AI) and floating-point operations per second (FLOP/s) performance.
To calculate these quantities, we need to collect the runtime, total FLOPs performed (the count, not the throughput per second), and the data movement (in bytes). 
The hierarchical Roofline models \cite{yang2019hierarchical} looks at data transactions between each pair of memory/cache levels, and on NVIDIA GPUs, we particularly focus on data transactions between these three levels, device memory (or HBM), L2 cache, and L1 cache. 

There is an \texttt{nvprof} based methodology for collecting Roofline data presented in this paper \cite{yang2019hierarchical}, and a more updated version based on Nsight Compute here, \cite{metho,yang2020howto}.
Nsight Compute is the NVIDIA performance tool that will replace \texttt{nvprof} in the future, and it has incorporated an HBM-only Roofline analysis feature in the CUDA 11 release.
In this paper, we will employ the methodology presented in \cite{metho,yang2020howto} for hierarchical Roofline analysis across three levels, HBM, L2, and L1, and particularly the arithmetic intensity (AI) and FLOP/s performance are calculated as follows.

\begin{align}
{\text{AI}}_{<\text{precision}>,<\text{level}>} & = \frac{
\text{\texttt{ncu} FLOPs}_{<\text{precision}>}
}{
\text{\texttt{ncu} Bytes}_{<\text{level}>}
}
\label{eqn:ai}\\
\text{FLOP/s}_{<\text{precision}>} & = \frac{
\text{\texttt{ncu} FLOPs}_{<\text{precision}>}
}{
\textit{\texttt{ncu} Runtime}
}\label{eqn:perf}
\end{align}
where {$<$\text{level}$>$} is L1, L2 and HBM, and {$<$\text{precision}$>$} is FP64 since the GPP kernel mostly performs double precision calculations.
The Nsight Compute based methodology in \cite{metho,yang2020howto} collects much fewer raw metrics and hardware counters than the one based on \texttt{nvprof}, in \cite{yang2019hierarchical}. It potentially presents lower overhead during profiling, and is thus a more recommended method.

\subsection{Machine Configuration\label{sec:machine}}

All the studies in this paper are conducted on the GPU partition of the Cori supercomputer at the National Energy Research Scientific Computing Center (NERSC), at Lawrence Berkeley National Laboratory (LBNL).
Cori GPU is primarily deployed for GPU porting, benchmarking, and testing efforts in the NERSC Exascale Science Applications Program (NESAP).
There are 18 GPU nodes in this partition, and each node contains two sockets of 20-core Intel Xeon Gold 6148 Skylake CPUs, 384 GB DDR4 memory, 930 GB on-node NVMe storage, and 8 NVIDIA V100 Volta GPUs.
Each of the GPUs has 80 Streaming Multiprocessors (SMs), 16 GB HBM2 memory, and is connected to the other seven in a `hybrid cube-mesh' topology.
Each SM on a Volta has 32 FP64 cores, and clocking at nominal freqency 1312 MHz, delivers for the entire GPU a theoretical peak of 6.7 TFLOP/s double precision performance.


\section{Optimization Journey}

In this section, we will discuss 8 steps we took to optimize the GPP kernel and improve its performance from 2.3 TFLOP/s to 3.7 TFLOP/s (double precision) on an NVIDIA V100 GPU.
Two benchmark systems, Si-214 and Si-510, are used to validate these optimizations, \texttt{v1} to \texttt{v8}, and the runtime and TFLOP/s performance for each step are listed in Tab.~\ref{tab:perf}.

\begin{table}[h]
\footnotesize
\centering
\setlength{\tabcolsep}{4.1pt}
\caption{\label{tab:perf}Optimization Path of the GPP Kernel on an NVIDIA V100 GPU}
\begin{tabular}{|c|c|c|c|c|c|c|}
\hline
\multirow{2}{*}{Version} &  \multicolumn{2}{c|}{Si-214} &  \multicolumn{2}{c|}{Si-510} & Warps  & Registers \\
\cline{2-5}
& Time & TFLOP/s & Time & TFLOP/s &  per SM  & per thread\\
\hline
v0 & 1.691 & \cellcolor{gray!25}2.337 & 24.705 & \cellcolor{gray!25}2.216 &  12 & 154 \\
\hline
v1 & 1.106 & \cellcolor{gray!25}2.629 & 13.269 & \cellcolor{gray!25}2.526 &  12 & 160 \\
\hline
v2 & 1.098 & \cellcolor{gray!25}2.628 & 13.260 & \cellcolor{gray!25}2.525 & 12 & 160 \\
\hline
v3 &  0.987 & \cellcolor{gray!25}2.647 & 11.983 & \cellcolor{gray!25}2.543 & 12 & 154 \\
\hline
v4 & 0.977 & \cellcolor{gray!25}2.754 & 11.246 & \cellcolor{gray!25}2.641 &  8 & 170 \\
\hline
v5 & 0.873 & \cellcolor{gray!25}2.901 & 10.257 & \cellcolor{gray!25}2.741 & 12 & 136 \\
\hline
v6 & 1.022 & \cellcolor{gray!25}2.392 & 11.923 & \cellcolor{gray!25}2.313 & 8 & 178 \\
\hline
v7 & 0.996 & \cellcolor{gray!25}2.548 & 10.901 & \cellcolor{gray!25}2.550 & 8 & 184 \\
\hline
v8 & 0.717 & \cellcolor{gray!25}3.710 & 7.565 & \cellcolor{gray!25}3.638 & 16 & 128 \\
\hline
\end{tabular}
\end{table}

The GPP kernel has a very high register usage throughout these optimization steps, and we have recorded its number of registers required per thread, and the actual number of active warps per SM, in Tab.~\ref{tab:perf}. Note that these numbers are the same for both benchmark systems, Si-214 and Si-510. 

This optimization process is a balancing act between register usage, SM occupancy, memory bandwidth usage, memory access pattern for different arrays, instruction latency, and arithmetic intensity, and we will discuss the 8 optimization steps in detail in the following subsections. 
The Roofline charts and Nsight Compute results in Fig.~1-8 are for the Si-214 benchmark, however, Si-510 presents a very similar profile, as can be seen by the trajectory in Tab.~\ref{tab:perf} (highlighted in gray).


\subsection*{v0. Baseline\label{sec:v0}}

\begin{figure}[t]
\centering
\makebox[0.5\textwidth][c]{\includegraphics[width=.5\textwidth]{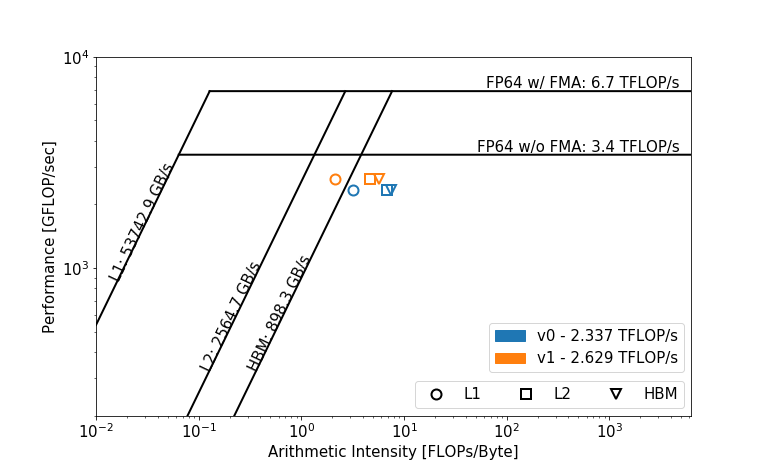}}
\caption{Hierarchical Roofline analysis of v0 and v1 for Si-214}
\label{fig:r0}
\end{figure}

The baseline version of the GPP kernel is a \texttt{collapse(3)} of the \texttt{band}, \texttt{igp}, and \texttt{ig} loops, and the \texttt{iw} loop is unrolled on each thread during the reduction. 
The reason behind this design for an initial CPU-to-GPU port is that it creates ample parallelism to fully utilize the available threads on a GPU.
This version of the kernel has a double precision performance of 2.3 TFLOP/s as shown in Fig.~\ref{fig:r0}, and there is little to no cache reuse between L1, L2 and HBM levels, as seen by the gaps between dots of the same color on the Roofline chart.

{\footnotesize
\begin{verbatim}
  !$acc parallel loop gang vector 
  !$acc reduction(+:...) collapse(3)
  do band = 1, nbands # O(1000) 
   do igp = 1, ngpown # O(1000)  
    do ig = 1, ncouls # O(10000)  
     do iw = 1, nw    # small, nw=2 
     ...
\end{verbatim}
}

\subsection*{v1. Replace divides\label{sec:v1}}

There are a few divide instructions in the GPP kernel, operating on complex double precision numbers, and these instructions present a very high execution latency, lower the warp issue rate, and can hurt our performance.
The division of two complex numbers ultimately end up with a floating point divide, e.g. \texttt{div.rn.f64}, and this instruction requires more cycles than a normal multiplication, addition, or FMA (fused multiplication and addition). 
NVIDIA GPUs have a faster, reciprocal instruction, e.g. \texttt{rcp.rn.f64}, and here we will try and coax the compiler to generate these reciprocals instead of divides.

\begin{figure}[h]
\centering
\makebox[0.5\textwidth][c]{\includegraphics[width=.5\textwidth]{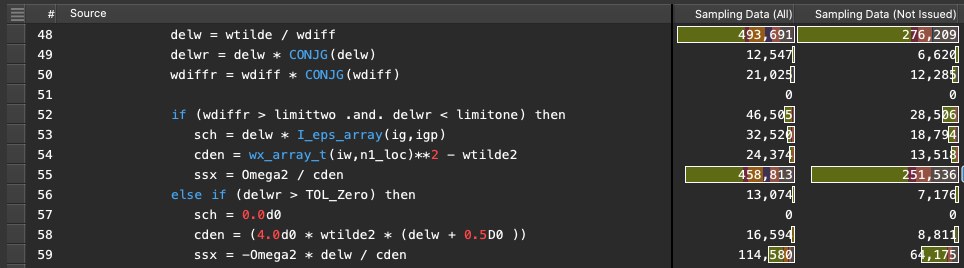}}

\vspace{0.05cm}
\makebox[0.5\textwidth][c]{\includegraphics[width=.5\textwidth]{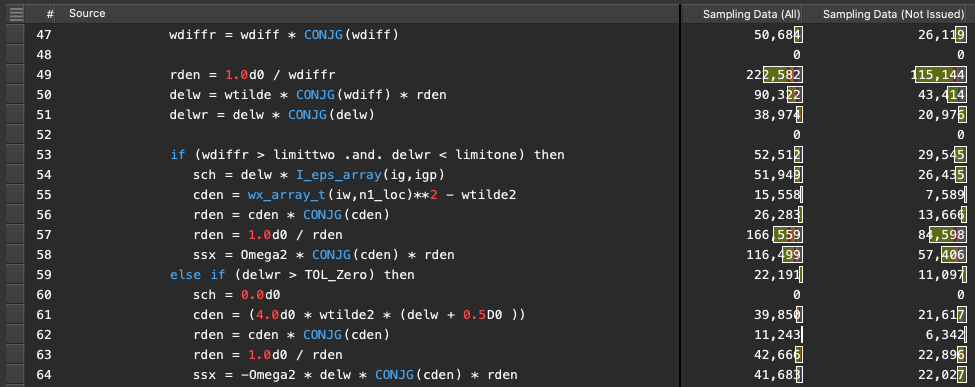}}
\caption{Replacing divides with reciprocals (Top: v0, Bottom: v1)}
\label{fig:v01}
\end{figure}

Fig.~\ref{fig:v01} shows the code difference before and after this optimization, and the Nsight Compute profile for the number of active warp samples that land on these lines of code. 
It is evident that the number has dropped significantly, and there is a 13\% improvement on the compute throughput, as shown in Fig.~\ref{fig:r0}.
The kernel also moved to be more bandwidth bound on the Roofline, which
is expected as more data needs to be drawn in order to satisfy the need of the faster computation.
We will observe this pattern again in \texttt{v3}.

Usually, performance dots on the Roofline chart move quite often as we optimize the code.
Even though the general narrative for Roofline based optimization process is to move these dots rightward (more compute bound) and upward (higher throughput, more performance), we do see these dots sometimes become more bandwidth bound, and even move downwards, as a temporary regression in performance.
However, as long as there is a clear philosophy and goal behind the design of these optimizations, performance gain can be redeemed with later optimizations. Steps \texttt{v6}, \texttt{v7} and \texttt{v8} will serve as a great example of this.

\subsection*{v2. Reduce branching\label{sec:v2}}

Even though GPU architectures have evolved to support branching more efficiently, it is still worth reducing the unnecessary branches to avoid thread divergence within a warp. 
In GPP, there is a 3-way branching code block shown below, and in this case, it can be further simplified to be 2-way as demonstrated in the \texttt{!After} section.
Even though this optimization did not bring any performance gain for GPP as shown in Fig.~\ref{fig:r1}, it has shown to be beneficial to other kernels in the full code BerkeleyGW, and is generally advisable.

{\footnotesize
\begin{verbatim}
  ! Before
  if (wdiffr > limittwo .and. delwr < limitone) then
    calculate sch, ssx 
  else if (delwr > TOL_Zero) then
    calculate sch, ssx 
  else
    sch = 0.0d0
    ssx = 0.0d0
  endif

  ! After
  sch = 0.0d0
  ssx = 0.0d0
  if (wdiffr > limittwo .and. delwr < limitone) then
    calculate sch, ssx 
  else if (delwr > TOL_Zero) then
    calculate sch, ssx
  endif
\end{verbatim}
}

\subsection*{v3. Replace \texttt{abs()}\label{sec:v3}}

\begin{figure}[h]
\centering
\makebox[0.5\textwidth][c]{\includegraphics[width=.5\textwidth]{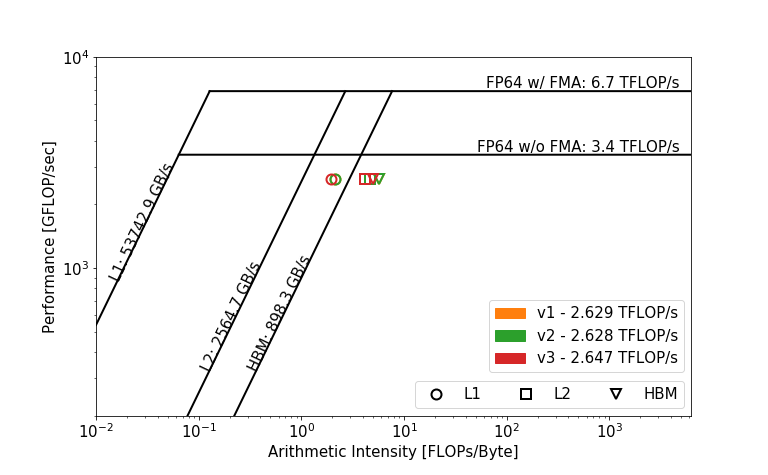}}
\caption{Hierarchical Roofline analysis of v1, v2 and v3 for Si-214}
\label{fig:r1}
\end{figure}
\begin{figure}[h]
\centering
\makebox[0.5\textwidth][c]{\includegraphics[width=.5\textwidth]{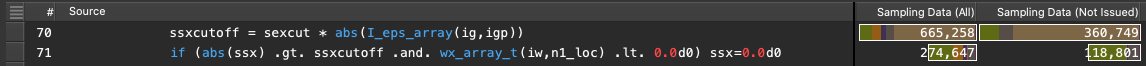}}

\vspace{0.05cm}
\makebox[0.5\textwidth][c]{\includegraphics[width=.5\textwidth]{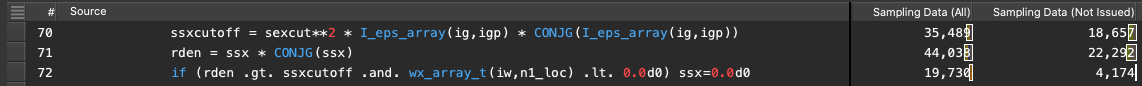}}
\caption{Replacing \texttt{abs()} with power 2 calculation (Top: v2, Bottom: v3)}
\label{fig:v23}
\end{figure}

As eluded to earlier in \texttt{v1}, \texttt{abs()} is also generally more expensive than a common multiply or add instruction, and in the case of GPP, \texttt{abs()} is only used for condition evaluation for \texttt{if/else} statements. 
Their explicit results are not needed, so we can try and replace them with power 2 calculations, i.e. $abs(a)*b<c$ is equivalent to $a*conj(a)*b^2<c^2$ for a complex number $a$.
This helps reduce the instruction latency and has proven to be beneficial in both the Nsight Compute profile in Fig.~\ref{fig:v23} and on the Roofline chart in Fig.~\ref{fig:r1}.

\subsection*{v4. Increase arithmetic intensity\label{sec:v4}}

Up to this point, the GPP kernel has been operating in an HBM bandwidth bound region on Roofline, with an arithmetic intensity just below the `machine balance' point (7.4 FLOPs/Byte on V100 for double precision and for the HBM level).
Since there is abundant parallelism and any two of the loops, \texttt{band}, \texttt{igp} and \texttt{ig}, can still provide enough parallelism to saturate a GPU, 
we can try collapse only two of them and leave the third running sequentially on each thread. 
This will increase the data reuse of certain arrays, increase the arithmetic intensity for the kernel, and move the kernel to a more compute bound region.
With a higher ceiling in the compute bound region (compared to the bandwidth bound region), we are more likely to reach higher performance if we can utilize the computational side of resources well.

However, the selection of the third loop needs to be done with care.
There are a few multi-dimensional arrays in the kernel, and serializing any of the loop indices could potentially break the memory coalescence for some of them.
Thorough investigation shows that \texttt{band} has the least number of arrays that use it as a non-first index, making it the logical choice.
With \texttt{band} unrolled, memory access for arrays such as \texttt{wtilde\_array(ig,igp)}, \texttt{I\_eps\_array(ig,igp)} and \texttt{aqsntemp(ig,band)} is still coalesced, because their fastest moving index \texttt{ig} is still mapped to the thread ID (note that Fortran is column major).
In OpenACC, this unrolling can be done as below.

{\footnotesize
\begin{verbatim}
  !$acc parallel loop gang vector 
  !$acc reduction(+:...) collapse(2)
  do igp = 1, ngpown    # O(1000)  
   do ig = 1, ncouls    # O(10000)  
    !$acc loop seq
    do band = 1, nbands # O(1000) 
     do iw = 1, nw      # small, nw=2 
     ...
\end{verbatim}
}

\begin{figure}[t]
\centering
\makebox[0.5\textwidth][c]{\includegraphics[width=.5\textwidth]{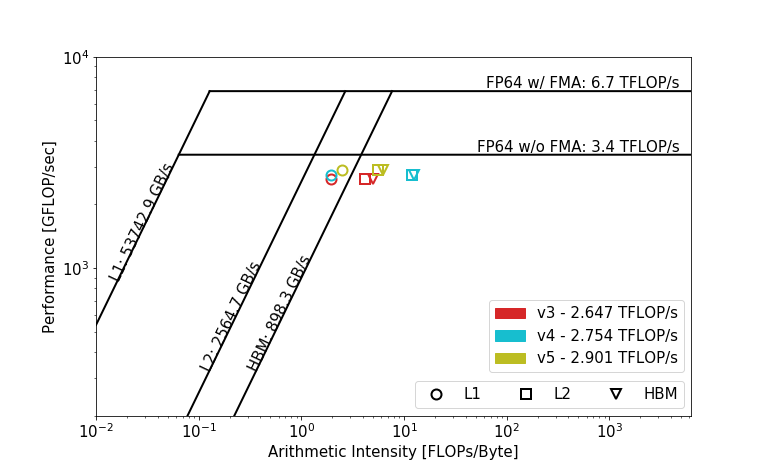}}
\caption{Hierarchical Roofline analysis of v3, v4 and v5 for Si-214}
\label{fig:r3}
\end{figure}

Thanks to this optimization, Fig.~\ref{fig:r3} shows a better separation between the L1 and L2 performance dots on Roofline (cyan), and the overall throughput has improved as well, from 2.6 TFLOP/s to 2.7 TFLOP/s.
The wider gaps between these dots suggest better cache reuse and data locality in L1, compared to previous versions, and with an arithmetic intensity of 12.5 FLOPs/Byte on the HBM level, it moves us back to the compute bound region, with more headroom to optimize for.


\subsection*{v5. Reduce branching, again\label{sec:v5}}


The baseline code \texttt{v0} unrolled the \texttt{iw} loop, and the reduction is done over four scalars due to the fact that OpenACC does not support array reductions at the time of writing. 
However, this constant branching creates warp divergence (especially when $nw>2$), increases register usage (which limits the number of active warps per SM), and requires more thread block synchronization as well.
In this step, we will move the \texttt{iw} loop outside the kernel, and reduce on a set of two scalars in a separate kernel.
This may result in redundant computation in both kernels ($nw=2$), however, the reduced register usage and increased occupancy (as seen in Tab.~\ref{tab:perf}) has proven to outweigh the duplication of certain calculations, as shown in Fig.~\ref{fig:r3}. 
The throughput has increased a little, and the kernel has moved to the left, back to the HBM bandwidth bound region.
We will attempt to tile the cache accesses to increase the arithmetic intensity in the next three steps, to move the kernel back to the compute bound region and to improve performance.


\subsection*{v6. Cache blocking\label{sec:v6}}

Cache blocking (or loop tiling) is a technique used to rearrange data access to pull subsets (blocks) of data into cache and to operate on this block to avoid repeatedly evicting it and fetching it back from more remote memory such as the main memory. 
This helps data reuse and gains cache locality.
In this step, we apply the blocking technique to \texttt{ig} and \texttt{band} loops, as shown in the pseudo code below.
Arrays with \texttt{ig} and \texttt{igp} indices can be shared as different \texttt{band} iterations are executed, and different threads can share the same \texttt{band} related arrays as well.

{\footnotesize
\begin{verbatim}
  ig_blksize = 128  
  band_blksize = 64  
  do iw = nstart, nend 
   !$ACC PARALLEL LOOP GANG VECTOR 
   !$ACC reduction(+:...) collapse(3)
   do band_blk = 1, band_blksize
    do igp = 1, ngpown  
     do ig_blk = 1, ig_blksize
      !$ACC LOOP SEQ
      do ig = ig_blk, ncouls, ig_blksize
       !$ACC LOOP SEQ
       do band = band_blk, nbands, band_blksize
       ...
\end{verbatim}
}

As a first attempt, this did not bring any performance benefit but has caused degradation in performance as shown in Fig.~\ref{fig:r5}.
However, the L1 and L2 cache locality has increased significantly, as seen by the wider gaps between L1 and L2 dots, and between L2 and HBM dots.
It has moved the kernel back to the compute bound region, and in the next two steps, we will try to make the kernel utilize the compute resources more efficiently, to move the dots upward on Roofline.

\subsection*{v7. Swap array index\label{sec:v7}}

\begin{figure}[h]
\centering
\makebox[0.5\textwidth][c]{\includegraphics[width=.5\textwidth]{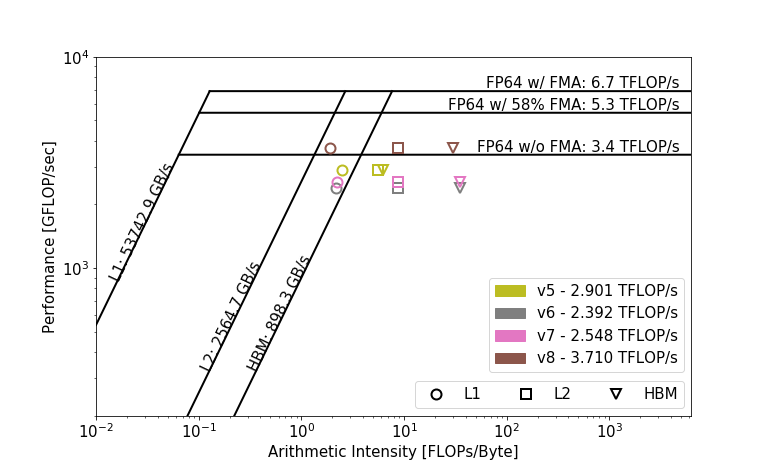}}
\caption{Hierarchical Roofline analysis of v5 to v8 for Si-214}
\label{fig:r5}
\end{figure}

To comply with the new memory access pattern after applying cache blocking, adjustment of the memory layout for certain arrays may be necessary.
For example, we need to swap the two indices of arrays \texttt{wx\_array\_t(iw,band)} and \texttt{aqsmtemp\_local(igp,band)}, to ensure that they are still accessed contiguously along the \texttt{band} index. 
This is different than ensuring memory coalescence among threads in a warp (which sometimes needs to be re-examined too), but it focuses on how these arrays are accessed between different iterations of \texttt{band} on the same thread.
As shown in Fig.~\ref{fig:r5}, this optimization has not provided much performance improvement, but it is necessary for the next optimization to take effect.

\subsection*{v8. Adjust thread block size\label{sec:v8}}

Currently, we are well in the compute bound region, and to improve the performance further, we should focus on the compute related resources, such as
whether we have enough concurrency on each SM.
As seen in Tab.~\ref{tab:perf}, the GPP kernel has had a very high register usage throughout this process.
In the case of \texttt{v7}, there are 184 registers required by each thread, and that has limited the number of active warps per SM to 8, which is only 1 thread block in this OpenACC kernel (the default \texttt{vector\_length} is 128 threads, i.e. 8 warps). 

This is a relatively low occupancy, considering that each SM can execute 2048 threads concurrently on a V100, which is 64 warps.
This low occupancy could reduce our latency-hiding ability on the GPU and be detrimental to the performance.
In the case of GPP, we already have a lot of long-latency instructions such as reciprocals - they are `long' compared to the regular multiplies and adds.
Also, the bulk of our operations are in double precision.
There are 64 FP32 cores, 64 Int32 cores and 32 FP64 cores on each SM, and this disparity in core count naturally dictates that FP64 instructions have longer execution latency 
than their counterpart FP32 or Int32 instructions. 
Switching to FP32 for self-energy calculation is not an option though, because of the high accuracy required, as in many other scientific applications.

However, to increase the occupancy, we can try manually limiting the number of registers per thread by either specifying \texttt{-Mcuda=maxregcount:128} to the PGI compiler, or by adding an OpenACC clause \texttt{vector\_length(512)} to the \texttt{parallel} construct in the code. For GPP, both cases will result in the same configuration of 512 threads per SM (i.e. 16 warps). 
This is double the number of warps we had before, and is a significant improvement in occupancy. 
However, the register spills created due to the squashing of the register usage can be a concern.

As we set a limit on the register usage, certain variables can not be stored in registers anymore, and they will instead be in the local memory, which is called a `register spill'. 
Local memory resides in device memory physically, and thus has a longer latency compared to the lower level caches (L1 and L2). 
Having said that, it is still worth a try to limit the register usage and gain better occupancy, because sometimes the benefit of the increased occupancy can outweigh the performance penality from register spills - just like in our case.
Fig.~\ref{fig:r5} shows that this optimization has improved GPP's performance significantly with a 3.7 TFLOP/s throughput now.
This of course is not possible if we did not apply cache blocking or the array index re-adjustment in \texttt{v6} and \texttt{v7}.

\begin{figure}[h]
\centering
\makebox[0.5\textwidth][c]{\includegraphics[width=.5\textwidth]{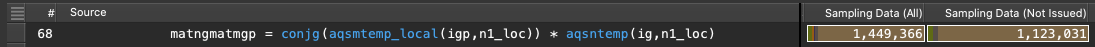}}

\vspace{0.02cm}
\makebox[0.5\textwidth][c]{\includegraphics[width=.5\textwidth]{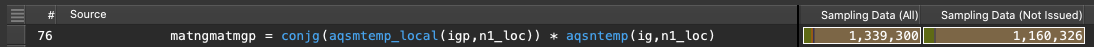}}

\vspace{0.03cm}
\makebox[0.5\textwidth][c]{\includegraphics[width=.5\textwidth]{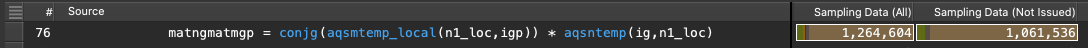}}

\vspace{0.03cm}
\makebox[0.5\textwidth][c]{\includegraphics[width=.5\textwidth]{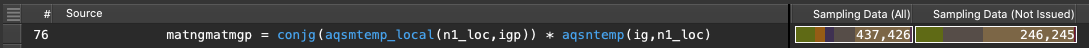}}
\caption{Nsight Compute analysis for the cache blocking steps (Top: v6, Middle: v7, and Bottom: v8)}
\label{fig:v68}
\end{figure}

\begin{figure}[h]
\centering
\makebox[0.5\textwidth][c]{\includegraphics[width=.5\textwidth]{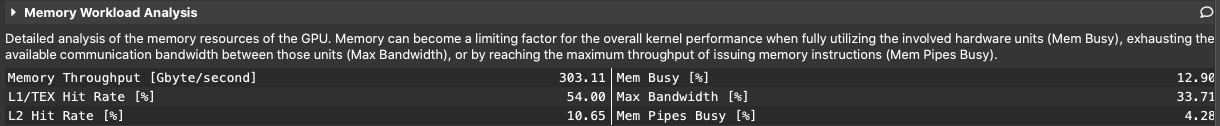}}

\vspace{0.05cm}
\makebox[0.5\textwidth][c]{\includegraphics[width=.5\textwidth]{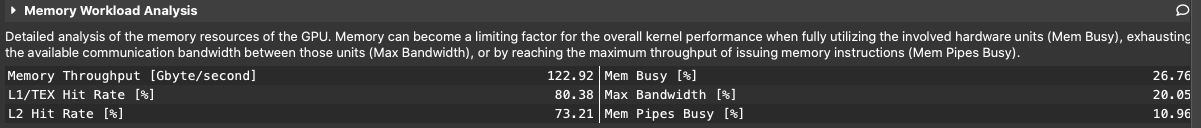}}
\caption{L1 and L2 cache hit rate (Top: v0, Bottom: v8)}
\label{fig:l2hit}
\end{figure}

The Nsight Compute profile validates the same conclusion in Fig.~\ref{fig:v68}, and as an example, the time spent on the \texttt{matngmatmgp} line has been largely reduced.
Throughout this entire process, from version \texttt{v0} to \texttt{v8}, the cache hit rate for both L1 and L2 has improved significantly as seen in Fig,~\ref{fig:l2hit}, which is a sign of successfully optimization as well.

Based on Nsight Compute metrics for \texttt{dmul}, \texttt{dadd} and \texttt{dfma} instructions, we have measured the ratio of the FMA instructions out of all FP64 instructions in GPP to 58\%, i.e. \texttt{dfma}/(\texttt{dmul}+\texttt{dadd}+ \texttt{dfma})=58\%.
Increasing the FMA ratio requires significant performance tuning effort and even if tuning on the lowest level, down to the assembly, we may still not get 100\% FMA ratio.
To realistically estimate the upper bound for the performance, we can customize our peak performance based on this 58\% FMA ratio  \cite{yang2019hierarchical}, i.e. with 58\% FMAs, we expect $(2\times0.58+1-0.58)/2=79\%$ of the theoretical FP64 peak at 1312 MHz, 6.7 TFLOP/s, and that yields the more realistic upper bound to be $0.79\times6.71 \text{ TFLOP/s}=5.3\text{ TFLOP/s}$.
It is the maximum attainable performance we can possibly achieve given the 58\% ratio, and with 3.71 TFLOP/s in \texttt{v8}, we have achieved 70\% of that ($3.71\text{ TFLOP/s}/5.3\text{ TFLOP/s}$), and 55\% of the theoretical peak ($3.71\text{ TFLOP/s}/6.71\text{ TFLOP/s}$), as shown in Fig.~\ref{fig:r5}.
This is significant contribution considering all the challenges we have faced, such as the complex double precision arithmetics, long-latency instructions such as divides and now reciprocals, complicated data access pattern for multiple multi-dimensional arrays, and the high register pressure and possible register spills.

\section{Conclusion}

This paper presents 8 optimization steps that have been taken to improve the performance of a Materials Science kernel GPP from 2.3 TFLOP/s to 3.7 TFLOP/s (double precision) on a single NVIDIA V100 GPU. 
An array of performance analysis and optimization techniques have been discussed in detail, including the use of the hierarchical Roofline performance model and the performance tool Nsight Compute.
The 3.7 TFLOP/s performance is at 55\% of the theoretical peak 6.7 TFLOP/s and 70\% of the kernel's customized attainable peak based on its FMA ratio, 5.3 TFLOP/s.
It is obtained given all the challenges faced such as complex number arithmetics, long-latency instructions, complicated data access pattern, and high register pressure.
These challenges are very common in many other scientific applications, and the techniques presented in this paper are expected to be of help to those with similar characteristics.


\bibliographystyle{IEEEtran}
\bibliography{IEEEabrv,references}

\end{document}